\documentclass[a4paper, 12pt, oneside]{Thesis}

\usepackage[T1]{fontenc}
\usepackage{verbatim}
\usepackage{vector}
\usepackage{lipsum}
\usepackage{tikz}
\usepackage{tkz-euclide}
\usepackage{hyperref}
\usepackage{subcaption}
\usepackage{environ}
\usepackage[noabbrev, capitalize, nameinlink]{cleveref}
\usepackage{pgfplots}
\usepackage{mathtools}
\usepackage[linesnumbered,ruled,vlined]{algorithm2e}
\usepackage{textcomp}
\usepackage{gensymb}

\usetikzlibrary{shapes}
\usetikzlibrary{arrows.meta}
\usetikzlibrary{shapes.geometric, arrows}
\tikzset{>={latex[width=1mm,length=2mm]}}
\hypersetup{urlcolor=blue, colorlinks=true,hidelinks}
\pgfplotsset{compat=1.14}

\def \linespaceing {1.5}
\DeclareMathOperator*{\XORsum}{\bigoplus}
\DeclareMathOperator*{\argmin}{arg\,min}
\newcommand{\hmm}[1]{\iffalse {1} \fi}

\DeclarePairedDelimiter\floor{\lfloor}{\rfloor}
\newlength{\myleftlen}
\newcommand{\setmyleftlen}[1]{\settowidth{\myleftlen}{\( \displaystyle
#1\)}}
\newcommand{\backup}{\hskip-\myleftlen\mkern-7mu}

\crefname{algocfline}{Algorithm}{Algorithms}

\definecolor{Paired-2}{RGB}{166,206,227}
\definecolor{Paired-1}{RGB}{31,120,180}
\definecolor{Paired-4}{RGB}{178,223,138}
\definecolor{Paired-3}{RGB}{51,160,44}
\definecolor{Paired-6}{RGB}{251,154,153}
\definecolor{Paired-5}{RGB}{227,26,28}
\definecolor{Paired-8}{RGB}{253,191,111}
\definecolor{Paired-7}{RGB}{255,127,0}
\definecolor{Paired-10}{RGB}{202,178,214}
\definecolor{Paired-9}{RGB}{106,61,154}
\definecolor{Paired-12}{RGB}{255,255,153}
\definecolor{Paired-13}{RGB}{0,0,0}
\definecolor{Paired-11}{RGB}{177,89,40}
\definecolor{Accent-1}{RGB}{127,201,127}
\definecolor{Accent-2}{RGB}{190,174,212}
\definecolor{Accent-3}{RGB}{253,192,134}
\definecolor{Accent-4}{RGB}{255,255,153}
\definecolor{Accent-5}{RGB}{56,108,176}
\definecolor{Accent-6}{RGB}{240,2,127}
\definecolor{Accent-7}{RGB}{191,91,23}
\definecolor{Accent-8}{RGB}{102,102,102}
\definecolor{Spectral-1}{RGB}{158,1,66}
\definecolor{Spectral-2}{RGB}{213,62,79}
\definecolor{Spectral-3}{RGB}{244,109,67}
\definecolor{Spectral-4}{RGB}{253,174,97}
\definecolor{Spectral-5}{RGB}{254,224,139}
\definecolor{Spectral-6}{RGB}{255,255,191}
\definecolor{Spectral-7}{RGB}{230,245,152}
\definecolor{Spectral-8}{RGB}{171,221,164}
\definecolor{Spectral-9}{RGB}{102,194,165}
\definecolor{Spectral-10}{RGB}{50,136,189}
\definecolor{Spectral-11}{RGB}{94,79,162}
\definecolor{Set1-1}{RGB}{228,26,28}
\definecolor{Set1-2}{RGB}{55,126,184}
\definecolor{Set1-3}{RGB}{77,175,74}
\definecolor{Set1-4}{RGB}{152,78,163}
\definecolor{Set1-5}{RGB}{255,127,0}
\definecolor{Set1-6}{RGB}{255,255,51}
\definecolor{Set1-7}{RGB}{166,86,40}
\definecolor{Set1-8}{RGB}{247,129,191}
\definecolor{Set1-9}{RGB}{153,153,153}
\definecolor{Set2-1}{RGB}{102,194,165}
\definecolor{Set2-2}{RGB}{252,141,98}
\definecolor{Set2-3}{RGB}{141,160,203}
\definecolor{Set2-4}{RGB}{231,138,195}
\definecolor{Set2-5}{RGB}{166,216,84}
\definecolor{Set2-6}{RGB}{255,217,47}
\definecolor{Set2-7}{RGB}{229,196,148}
\definecolor{Set2-8}{RGB}{179,179,179}
\definecolor{Dark2-1}{RGB}{27,158,119}
\definecolor{Dark2-2}{RGB}{217,95,2}
\definecolor{Dark2-3}{RGB}{117,112,179}
\definecolor{Dark2-4}{RGB}{231,41,138}
\definecolor{Dark2-5}{RGB}{102,166,30}
\definecolor{Dark2-6}{RGB}{230,171,2}
\definecolor{Dark2-7}{RGB}{166,118,29}
\definecolor{Dark2-8}{RGB}{102,102,102}
\definecolor{Reds-1}{RGB}{255,245,240}
\definecolor{Reds-2}{RGB}{254,224,210}
\definecolor{Reds-3}{RGB}{252,187,161}
\definecolor{Reds-4}{RGB}{252,146,114}
\definecolor{Reds-5}{RGB}{251,106,74}
\definecolor{Reds-6}{RGB}{239,59,44}
\definecolor{Reds-7}{RGB}{203,24,29}
\definecolor{Reds-8}{RGB}{165,15,21}
\definecolor{Reds-9}{RGB}{103,0,13}
\definecolor{Greens-1}{RGB}{247,252,245}
\definecolor{Greens-2}{RGB}{229,245,224}
\definecolor{Greens-3}{RGB}{199,233,192}
\definecolor{Greens-4}{RGB}{161,217,155}
\definecolor{Greens-5}{RGB}{116,196,118}
\definecolor{Greens-6}{RGB}{65,171,93}
\definecolor{Greens-7}{RGB}{35,139,69}
\definecolor{Greens-8}{RGB}{0,109,44}
\definecolor{Greens-9}{RGB}{0,68,27}
\definecolor{Blues-1}{RGB}{247,251,255}
\definecolor{Blues-2}{RGB}{222,235,247}
\definecolor{Blues-3}{RGB}{198,219,239}
\definecolor{Blues-4}{RGB}{158,202,225}
\definecolor{Blues-5}{RGB}{107,174,214}
\definecolor{Blues-6}{RGB}{66,146,198}
\definecolor{Blues-7}{RGB}{33,113,181}
\definecolor{Blues-8}{RGB}{8,81,156}
\definecolor{Blues-9}{RGB}{8,48,107}
\definecolor{grey_l}{RGB}{222,222,222}
\definecolor{Paired-1l}{RGB}{236,251,252}
\definecolor{Paired-3l}{RGB}{230,255,178}
\definecolor{Paired-5l}{RGB}{255,206,205}

\DeclareTextFontCommand{\variable}{\fontfamily{phv}\selectfont\itshape}
\DeclareTextFontCommand{\datastruct}{\fontfamily{phv}\selectfont\bfseries}
\DeclareTextFontCommand{\function}{\fontfamily{cmr}\selectfont}

\begin{document}

\selectlanguage{english}

\frontmatter

\title  {Towards Practical Software Stack Decoding of Polar Codes}
\authors  {Harsh Aurora, Warren Gross}
\maketitle

\setstretch{\linespaceing}

\cfoot{\thepage}

\mainmatter
\pagestyle{plain}

\section{Introduction} \label{s1}

	Polar codes were proposed by Erdal Arikan in 2008 \cite{2} as the first set of linear block codes that have an explicit construction and provably achieve the symmetric capacity of a binary memoryless channel. Polar codes faced initial resistance due to the low throughput of the sequential successive cancellation (SC) decoding algorithm, as well as its mediocre error-correction performance at short code lengths. The successive cancellation list (SCL) algorithm \cite{26} and its CRC-aided variant \cite{5} enabled optimal decoding performance at short lengths, and fast simplified successive cancellation decoding \cite{4,7,8} improved the throughput performance considerably, thereby deeming polar codes a viable candidate for practical applications. In 2016, polar codes were selected by 3GPP as one of the error-correcting codes to be used in the enhanced Mobile Broadband (eMBB) control channel \cite{e1,e2}. 

	The successive cancellation stack (SCS) algorithm proposed in 2012 \cite{10,11} provides similar error-correcting performance as the SCL algorithm with a complexity that varies with the channel conditions. At high channel noise, SCS has the same complexity as SCL, and as channel noise decreases, the SCS complexity approaches that of SC, making it an attractive candidate. Although the SCS algorithm has an attractive complexity, software implementations report a mediocre T/P performance. This technical report outlines a method to apply the fast simplified decoding scheme in \cite{4,7} to the reduced memory stack decoder in \cite{22}, resulting in a software T/P improvement of over two orders of magnitude, from 9 Kbps to 930 Kbps.

	This report is organized as follows: \Cref{s2} provides relevant background information regarding polar codes and the pertinent decoding algorithms. \Cref{s3} highlights key software implementation details of the decoders. \Cref{s4} describes the fast simplified scheme applied to stack decoding, and \cref{s5} presents and discusses the simulation results. Finally, \cref{s6} concludes this report.

\section{Background} \label{s2}
	\subsection{Polar codes} \label{s2-1}
		Polar codes asymptotically achieve the symmetric channel capacity for a B-DMC $W$ by considering a set of N independent copies of $W$ and recursively applying a polarizing transform $F = \left[\begin{smallmatrix} 1&0\\ 1&1 \end{smallmatrix} \right]$ to the inputs of the channels, resulting second set of N channels $\left\{W^{(i)}_{N}\right\}$ that are said to be polarized in the sense that $K$ of the inputs are completely reliable, while the remaining $(N-K)$ inputs are completely unreliable, and as $N \rightarrow \infty$ the fraction $\frac{K}{N} \rightarrow I(W)$.

					A polar code of length $N$ and message bit length $K$ shall be denoted by $PC(N,K)$. Given an information bit set $\mathcal{A}$ of size $K$ and a corresponding frozen bit set $\mathcal{A^{C}}$ of size $(N-K)$, the input $u_{0}^{N-1}$ to the polarized channels is constructed from a message bit sequence $m_{0}^{K-1}$ by placing the bits at the indices contained in a $\mathcal{A}$, and setting the remaining indices to 0. The encoding step to generate the codeword $c_{0}^{N-1}$ can then be expressed as the matrix multiplication

\begin{figure}[t]
				\centering
				\begin{tikzpicture}[every node/.style={scale=0.8}]
					\tikzset{
						chan/.style={rectangle,minimum height=1 cm,minimum width=1.5 cm,draw},
						xor/.style={circle,draw},
						l/.style={Paired-11},
						p/.style={Spectral-10},
					}

					\draw[l,thick,dashed] (1.35,-1) -- (1.35, 8.7);
					\draw[l,thick,dashed] (4.05,-1) -- (4.05, 8.7);
					\draw[l,thick,dashed] (7.45,-1) -- (7.45, 8.7);

					\draw[p,thick,->] (-1,8) -- (-1, -1);
					\draw[l,thick,->] (0,9) -- (12,9);

					\node[anchor=west, scale=1.3,l] at (12,9) {$\boldsymbol{\lambda}$};
					\node[anchor=north, scale=1.3,p] at (-1,-1) {$\boldsymbol{\phi}$};

					\node[p, scale=1.3] at (10.9,3.5) {\textbf{0}};

					\node[p, scale=1.3] at (6.1,5.5) {\textbf{0}};
					\node[p, scale=1.3] at (6.1,1.5) {\textbf{1}};

					\node[p, scale=1.3] at (2.7,6.5) {\textbf{0}};
					\node[p, scale=1.3] at (2.7,4.5) {\textbf{1}};
					\node[p, scale=1.3] at (2.7,2.5) {\textbf{2}};
					\node[p, scale=1.3] at (2.7,0.5) {\textbf{3}};

					\node[p, scale=1.3] at (0.2,7) {\textbf{0}};
					\node[p, scale=1.3] at (0.2,6) {\textbf{1}};
					\node[p, scale=1.3] at (0.2,5) {\textbf{2}};
					\node[p, scale=1.3] at (0.2,4) {\textbf{3}};
					\node[p, scale=1.3] at (0.2,3) {\textbf{4}};
					\node[p, scale=1.3] at (0.2,2) {\textbf{5}};
					\node[p, scale=1.3] at (0.2,1) {\textbf{6}};
					\node[p, scale=1.3] at (0.2,0) {\textbf{7}};

					\node[l, scale=1.3] at (10.2,8.5) {\textbf{3}};
					\node[l, scale=1.3] at (5.8,8.5) {\textbf{2}};
					\node[l, scale=1.3] at (2.8,8.5) {\textbf{1}};
					\node[l, scale=1.3] at (0.4,8.5) {\textbf{0}};

					\node[anchor=east, p, thick, rectangle, dashed, draw, minimum height=1.0 cm,minimum width=0.7 cm] at (0,0) { };
					\node[anchor=east, p, thick, rectangle, dashed, draw, minimum height=1.0 cm,minimum width=0.7 cm] at (0,1) { };
					\node[anchor=east, p, thick, rectangle, dashed, draw, minimum height=1.0 cm,minimum width=0.7 cm] at (0,2) { };
					\node[anchor=east, p, thick, rectangle, dashed, draw, minimum height=1.0 cm,minimum width=0.7 cm] at (0,3) { };
					\node[anchor=east, p, thick, rectangle, dashed, draw, minimum height=1.0 cm,minimum width=0.7 cm] at (0,4) { };
					\node[anchor=east, p, thick, rectangle, dashed, draw, minimum height=1.0 cm,minimum width=0.7 cm] at (0,5) { };
					\node[anchor=east, p, thick, rectangle, dashed, draw, minimum height=1.0 cm,minimum width=0.7 cm] at (0,6) { };
					\node[anchor=east, p, thick, rectangle, dashed, draw, minimum height=1.0 cm,minimum width=0.7 cm] at (0,7) { };

					\node[p, thick, rectangle, dashed, draw, minimum height=2.3 cm,minimum width=1 cm] at (2,0.5) { };
					\node[p, thick, rectangle, dashed, draw, minimum height=2.3 cm,minimum width=1 cm] at (2,2.5) { };
					\node[p, thick, rectangle, dashed, draw, minimum height=2.3 cm,minimum width=1 cm] at (2,4.5) { };
					\node[p, thick, rectangle, dashed, draw, minimum height=2.3 cm,minimum width=1 cm] at (2,6.5) { };

					\node[p, thick, rectangle, dashed, draw, minimum height=4.8 cm,minimum width=1.9 cm] at (5.05,1.5) { };
					\node[p, thick, rectangle, dashed, draw, minimum height=4.8 cm,minimum width=1.9 cm] at (5.05,5.5) { };

					\node[p, thick, rectangle, dashed, draw, minimum height=9.9 cm,minimum width=3.7 cm] at (9.15,3.5) { };

					\tkzDefPoint(0    , 7   ){u0}
					\tkzDefPoint(0    , 6   ){u1}
					\tkzDefPoint(0    , 5   ){u2}
					\tkzDefPoint(0    , 4   ){u3}
					\tkzDefPoint(0    , 3   ){u4}
					\tkzDefPoint(0    , 2   ){u5}
					\tkzDefPoint(0    , 1   ){u6}
					\tkzDefPoint(0    , 0   ){u7}

					\tkzDefPoint(2    , 7   ){s00}
					\tkzDefPoint(2    , 6   ){s01}
					\tkzDefPoint(2    , 5   ){s02}
					\tkzDefPoint(2    , 4   ){s03}
					\tkzDefPoint(2    , 3   ){s04}
					\tkzDefPoint(2    , 2   ){s05}
					\tkzDefPoint(2    , 1   ){s06}
					\tkzDefPoint(2    , 0   ){s07}

					\tkzDefPoint(4.7  , 7   ){s10}
					\tkzDefPoint(5.4  , 6   ){s11}
					\tkzDefPoint(4.7  , 5   ){s12}
					\tkzDefPoint(5.4  , 4   ){s13}
					\tkzDefPoint(4.7  , 3   ){s14}
					\tkzDefPoint(5.4  , 2   ){s15}
					\tkzDefPoint(4.7  , 1   ){s16}
					\tkzDefPoint(5.4  , 0   ){s17}

					\tkzDefPoint(8.1  , 7   ){s20}
					\tkzDefPoint(8.8  , 6   ){s21}
					\tkzDefPoint(9.5  , 5   ){s22}
					\tkzDefPoint(10.2 , 4   ){s23}
					\tkzDefPoint(8.1  , 3   ){s24}
					\tkzDefPoint(8.8  , 2   ){s25}
					\tkzDefPoint(9.5  , 1   ){s26}
					\tkzDefPoint(10.2 , 0   ){s27}

					\tkzDefPoint(12.9 , 7   ){c0}
					\tkzDefPoint(12.9 , 6   ){c1}
					\tkzDefPoint(12.9 , 5   ){c2}
					\tkzDefPoint(12.9 , 4   ){c3}
					\tkzDefPoint(12.9 , 3   ){c4}
					\tkzDefPoint(12.9 , 2   ){c5}
					\tkzDefPoint(12.9 , 1   ){c6}
					\tkzDefPoint(12.9 , 0   ){c7}

					\node[anchor=east] (u0) at (u0) {$0$};
					\node[anchor=east] (u1) at (u1) {$0$};
					\node[anchor=east] (u2) at (u2) {$0$};
					\node[anchor=east] (u3) at (u3) {$m_0$};
					\node[anchor=east] (u4) at (u4) {$0$};
					\node[anchor=east] (u5) at (u5) {$m_1$};
					\node[anchor=east] (u6) at (u6) {$m_2$};
					\node[anchor=east] (u7) at (u7) {$m_3$};

					\node[xor] (s00) at (s00) {\textbf{+}};
					\node[xor] (s02) at (s02) {\textbf{+}};
					\node[xor] (s04) at (s04) {\textbf{+}};
					\node[xor] (s06) at (s06) {\textbf{+}};

					\node[xor] (s10) at (s10) {\textbf{+}};
					\node[xor] (s11) at (s11) {\textbf{+}};
					\node[xor] (s14) at (s14) {\textbf{+}};
					\node[xor] (s15) at (s15) {\textbf{+}};

					\node[xor] (s20) at (s20) {\textbf{+}};
					\node[xor] (s21) at (s21) {\textbf{+}};
					\node[xor] (s22) at (s22) {\textbf{+}};
					\node[xor] (s23) at (s23) {\textbf{+}};

					\node[anchor=west] (c0) at (c0) {$c_0$};
					\node[anchor=west] (c1) at (c1) {$c_1$};
					\node[anchor=west] (c2) at (c2) {$c_2$};
					\node[anchor=west] (c3) at (c3) {$c_3$};
					\node[anchor=west] (c4) at (c4) {$c_4$};
					\node[anchor=west] (c5) at (c5) {$c_5$};
					\node[anchor=west] (c6) at (c6) {$c_6$};
					\node[anchor=west] (c7) at (c7) {$c_7$};

					\draw[->] (u0)  -- (s00);
					\draw[->] (s00) -- (s10);
					\draw[->] (s10) -- (s20);
					\draw[->] (s20) -- (c0) ;

					\draw[- ] (u1)  -- (s01);
					\draw[->] (s01) -- (s11);
					\draw[->] (s11) -- (s21);
					\draw[->] (s21) -- (c1) ;

					\draw[->] (u2)  -- (s02);
					\draw[- ] (s02) -- (s12);
					\draw[->] (s12) -- (s22);
					\draw[->] (s22) -- (c2) ;

					\draw[- ] (u3)  -- (s03);
					\draw[- ] (s03) -- (s13);
					\draw[->] (s13) -- (s23);
					\draw[->] (s23) -- (c3) ;

					\draw[->] (u4)  -- (s04);
					\draw[->] (s04) -- (s14);
					\draw[- ] (s14) -- (s24);
					\draw[->] (s24) -- (c4) ;

					\draw[- ] (u5)  -- (s05);
					\draw[->] (s05) -- (s15);
					\draw[- ] (s15) -- (s25);
					\draw[->] (s25) -- (c5) ;

					\draw[->] (u6)  -- (s06);
					\draw[- ] (s06) -- (s16);
					\draw[- ] (s16) -- (s26);
					\draw[->] (s26) -- (c6) ;

					\draw[- ] (u7)  -- (s07);
					\draw[- ] (s07) -- (s17);
					\draw[- ] (s17) -- (s27);
					\draw[->] (s27) -- (c7) ;

					\draw[<-] (s00) -- (s01);
					\draw[<-] (s02) -- (s03);
					\draw[<-] (s04) -- (s05);
					\draw[<-] (s06) -- (s07);

					\draw[<-] (s10) -- (s12);
					\draw[<-] (s11) -- (s13);
					\draw[<-] (s14) -- (s16);
					\draw[<-] (s15) -- (s17);

					\draw[<-] (s20) -- (s24);
					\draw[<-] (s21) -- (s25);
					\draw[<-] (s22) -- (s26);
					\draw[<-] (s23) -- (s27);

				\end{tikzpicture}
			\caption{Encoding tree for $PC(8,4)$ with $\mathcal{A} = \{3,5,6,7\}$.}	
			\label{fig:encoder}
			\end{figure}
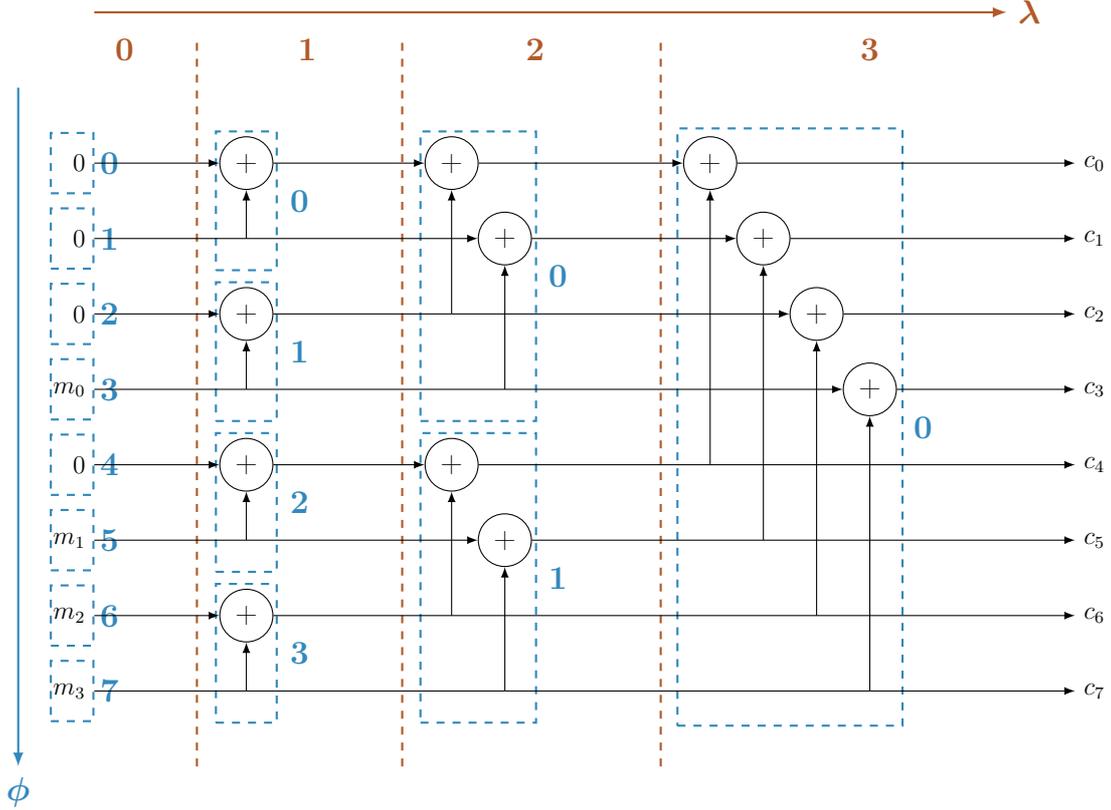

			\begin{center}
			$c_{0}^{N-1} = u_{0}^{N-1}F^{\otimes n}\,\,,$
			\end{center}

			where $n=log_2N$ and $F^{\otimes n}$ is the $n^{th}$ Kronecker power of the kernel $F$, and can be represented by the XOR tree shown in \cref{fig:encoder}. The tree has $n$ stages, and the variable $\lambda \in [0,n]$ is used to denote the current stage in the tree. Given a stage $\lambda$, there are ($n-\lambda+1$) branches denoted by $\phi \in [0,(n-\lambda)]$, and the size of each branch is $\Lambda=2^\lambda$.

	\subsection{Decoding algorithms} \label{s2-2}
		All decoding algorithms in this section are described in the LLR domain.

		\subsubsection{Successive cancellation}

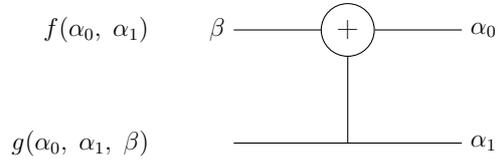
\begin{figure}[t]
				\centering
				\begin{tikzpicture}[every node/.style={scale=0.8}]
					\tikzset{
						chan/.style={rectangle,minimum height=1 cm,minimum width=1.5 cm,draw},
						xor/.style={circle,draw},
					}
					\node[anchor=east] (f) at (0,1.5) {$f(\alpha_0,\;\alpha_1)$};
					\node[anchor=east] (g) at (0,0) {$g(\alpha_0,\;\alpha_1,\;\beta)$};

					\node[anchor=east] (u0) at (1,1.5) {$\beta$};			
					\node[anchor=east] (u1) at (1,0) {};

					\node[xor] (xor) at (2.5,1.5) {\textbf{+}};
					
					\node[anchor=west] (y0) at (4,1.5) {$\alpha_0$};			
					\node[anchor=west] (y1) at (4,0) {$\alpha_1$};
					
					\draw (u1) -- (y1);
					\draw (u0) -- (xor);
					\draw (xor) -- (y0);

					\draw (2.5,0) -- (xor);
				\end{tikzpicture}
			\caption{Min-sum approximation over the polarizing kernel.}
			\label{fig:minsum}
			\end{figure}

			The successive cancellation (SC) decoding algorithm \cite{2} operates on the encoding tree, propagating that channel values $LLR(y_i)$ from stage $n$ to produce $LLR\left(y_0^{N-1}, \hat{u}_0^{i-1}| \hat{u}^i\right)$ at stage 0, according to the min-sum approximation \cite{15} in \cref{fig:minsum,eq:minsum}. The estimate $\hat{u}_i$ can then be made following \cref{eq:scest}.

			\begin{align}
				f(\alpha_0,\;\alpha_1) &= sign(\alpha_0)\;sign(\alpha_1)\;min(|\alpha_0|,|\alpha_1|) & ,\,\, & \alpha \in \mathbb{R}\\
				g(\alpha_0,\;\alpha_1,\beta) &= \alpha_0 + (-1)^\beta\alpha_1 & ,\,\, & \alpha \in \mathbb{R},\,\beta \in [0,1]
			\label{eq:minsum}
			\end{align}

			\begin{equation}
				\hat{u}_i = 
					\begin{cases}
						HD\left(LLR\left(y_0^{N-1}, \hat{u}_0^{i-1}| \hat{u}^i\right)\right) & \quad \text{if } i \in \mathcal{A}\\
						0 & \quad \text{otherwise}\\
					\end{cases}
			\label{eq:scest}
			\end{equation}

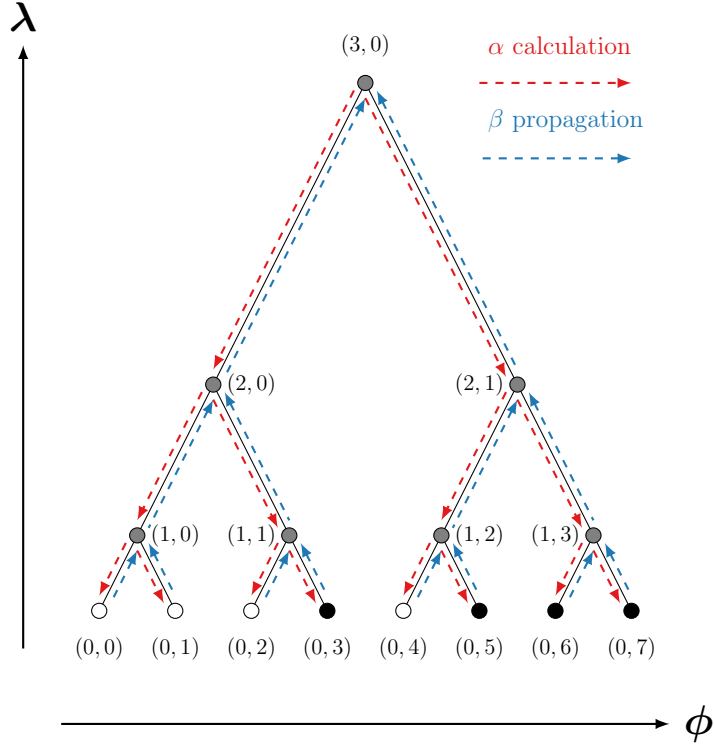
\begin{figure}[t]
				\centering
				\begin{tikzpicture}
					\tikzset{
						alphaline/.style={Paired-5, dashed, thick, ->},
						betaline/.style={Paired-1, dashed, thick, <-},
			    	Lscale/.style={scale=0.7},
			    	l/.style={Paired-11},
						p/.style={Spectral-10},
					}

					\draw[thick,->] (-1,-0.5) -- (-1,7.5);
					\draw[thick,->] (-0.5,-1.5) -- (7.5,-1.5);

					\node[anchor=south, scale=1.3] at (-1,7.5) {$\boldsymbol{\lambda}$};
					\node[anchor=west, scale=1.3] at (7.5,-1.5) {$\boldsymbol{\phi}$};

					\node[scale=0.8,anchor=west,Paired-5] at (5,7.5) {$\alpha$ calculation};
					\draw[alphaline] (5,7) -- (7,7);
					\node[scale=0.8,anchor=west,Paired-1] at (5,6.5) {$\beta$ propagation};
					\draw[betaline] (7,6) -- (5,6);

					\draw (3.5,7) -- (1.5,3);
					\draw[alphaline] (3.35,6.9) -- (1.5,3.2);
					\draw[betaline]  (3.5,6.8) -- (1.65,3.1);
					\draw (3.5,7) -- (5.5,3);
					\draw[alphaline] (3.5,6.8) -- (5.35,3.1);		
					\draw[betaline]  (3.65,6.9) -- (5.5,3.2);

					\draw (1.5,3) -- (0.5,1);
					\draw[alphaline] (1.35,2.9) -- (0.5,1.2);
					\draw[betaline] (1.5,2.8) -- (0.65,1.1);
					\draw (1.5,3) -- (2.5,1);
					\draw[alphaline] (1.5,2.8) -- (2.35,1.1);
					\draw[betaline] (1.65,2.9) -- (2.5,1.2);	
					\draw (5.5,3) -- (4.5,1);
					\draw[alphaline] (5.35,2.9) -- (4.5,1.2);
					\draw[betaline] (5.5,2.8) -- (4.65,1.1);
					\draw (5.5,3) -- (6.5,1);
					\draw[alphaline] (5.5,2.8) -- (6.35,1.1);
					\draw[betaline] (5.65,2.9) -- (6.5,1.2);

					\draw (0.5,1) -- (0,0);
					\draw[alphaline] (0.35,0.9) -- (0,0.2);
					\draw[betaline] (0.5,0.8) -- (0.15,0.1);
					\draw (0.5,1) -- (1,0);
					\draw[alphaline] (0.5,0.8) -- (0.85,0.1);
					\draw[betaline] (0.65,0.9) -- (1,0.2);
					\draw (2.5,1) -- (2,0);
					\draw[alphaline] (2.35,0.9) -- (2,0.2);
					\draw[betaline] (2.5,0.8) -- (2.15,0.1);
					\draw (2.5,1) -- (3,0);
					\draw[alphaline] (2.5,0.8) -- (2.85,0.1);
					\draw[betaline] (2.65,0.9) -- (3,0.2);
					\draw (4.5,1) -- (4,0);
					\draw[alphaline] (4.35,0.9) -- (4,0.2);
					\draw[betaline] (4.5,0.8) -- (4.15,0.1);
					\draw (4.5,1) -- (5,0);
					\draw[alphaline] (4.5,0.8) -- (4.85,0.1);
					\draw[betaline] (4.65,0.9) -- (5,0.2);
					\draw (6.5,1) -- (6,0);
					\draw[alphaline] (6.35,0.9) -- (6,0.2);
					\draw[betaline] (6.5,0.8) -- (6.15,0.1);
					\draw (6.5,1) -- (7,0);
					\draw[alphaline] (6.5,0.8) -- (6.85,0.1);
					\draw[betaline] (6.65,0.9) -- (7,0.2);

					\fill[white] (0,0) circle (0.1);
					\draw (0,0) circle (0.1);
					\draw (0,-0.5) node[Lscale] {$(0, 0)$};
					\fill[white] (1,0) circle (0.1);
					\draw (1,0) circle (0.1);
					\draw (1,-0.5) node[Lscale] {$(0, 1)$};
					\fill[white] (2,0) circle (0.1);
					\draw (2,0) circle (0.1);
					\draw (2,-0.5) node[Lscale] {$(0, 2)$};
					\fill[black] (3,0) circle (0.1);
					\draw (3,0) circle (0.1);
					\draw (3,-0.5) node[Lscale] {$(0, 3)$};
					\fill[white] (4,0) circle (0.1);
					\draw (4,0) circle (0.1);
					\draw (4,-0.5) node[Lscale] {$(0, 4)$};
					\fill[black] (5,0) circle (0.1);
					\draw (5,0) circle (0.1);
					\draw (5,-0.5) node[Lscale] {$(0, 5)$};
					\fill[black] (6,0) circle (0.1);
					\draw (6,0) circle (0.1);
					\draw (6,-0.5) node[Lscale] {$(0, 6)$};
					\fill[black] (7,0) circle (0.1);
					\draw (7,0) circle (0.1);
					\draw (7,-0.5) node[Lscale] {$(0, 7)$};

					\fill[gray] (0.5,1) circle (0.1);
					\draw (0.5,1) circle (0.1);
					\draw (1,1) node[Lscale] {$(1, 0)$};
					\fill[gray] (2.5,1) circle (0.1);
					\draw (2.5,1) circle (0.1);
					\draw (2,1) node[Lscale] {$(1, 1)$};
					\fill[gray] (4.5,1) circle (0.1);
					\draw (4.5,1) circle (0.1);
					\draw (5,1) node[Lscale] {$(1, 2)$};
					\fill[gray] (6.5,1) circle (0.1);
					\draw (6.5,1) circle (0.1);
					\draw (6,1) node[Lscale] {$(1, 3)$};

					\fill[gray] (1.5,3) circle (0.1);
					\draw (1.5,3) circle (0.1);
					\draw (2,3) node[Lscale] {$(2, 0)$};
					\fill[gray] (5.5,3) circle (0.1);
					\draw (5.5,3) circle (0.1);
					\draw (5,3) node[Lscale] {$(2, 1)$};

					\fill[gray] (3.5,7) circle (0.1);
					\draw (3.5,7) circle (0.1);
					\draw (3.5,7.5) node[Lscale] {$(3,0)$};
				\end{tikzpicture}
			\caption{SC decoding tree and schedule for $PC(8,4)$.}
			\label{fig:scdectree}
			\end{figure}

			The XOR encoding tree in \cref{fig:encoder} is reinterpreted as a binary tree as shown in \cref{fig:scdectree}, and the stage $\lambda$ and branch $\phi$ is used to identify each node, denoted by $(\lambda,\,\phi)$. A node $v = (\lambda,\,\phi)$ has associated LLR values $\alpha_v[i]$ and bit estimates $\beta_v[i]$, where $i \in [0,\Lambda-1]$. The LLR's of the root node at $(n,0)$ are obtained directly from the channel output, and the LLR's of child nodes $v = (\lambda,\,\phi)$ are calculated from the parent node $p = \left(\lambda+1,\,\floor*{\frac{\phi}{2}}\right)$ and previous branch $u = (\lambda,\,\phi-1)$ according to \cref{eq:llrcalc}. The LLR $\alpha_{(0,i)}$ calculated for a leaf node is the desired $LLR\left(y_0^{N-1}, \hat{u}_0^{i-1}| \hat{u}^i\right)$.

			\begin{align}
				\begin{aligned}
					\alpha_{(n,0)}[i] &= LLR(y[i]) & ,\,\, & i \in [0,N-1]\\\\
					\alpha_v[i] &=
						\begin{cases}
							f(\alpha_p[i],\,\alpha_p[i+\Lambda]) & \quad \text{if $\phi$ is even}\\
							g(\alpha_p[i],\,\alpha_p[i+\Lambda],\,\beta_u[i]) & \quad \text{if $\phi$ is odd}\\
						\end{cases}& ,\,\, & i \in [0,\Lambda-1]\\
				\end{aligned}
			\label{eq:llrcalc}
			\end{align}

			The bit estimates of the leaf nodes at $(0,i)$ correspond to $\hat{u}_i$, and are obtained via a hard decision on its LLR. The bit estimates of parent nodes $v = (\lambda,\,\phi)$ are calculated by propagating those of both the child nodes $l = (\lambda-1,\,2\phi)$ and $r = (\lambda-1,\,2\phi+1)$, as shown in \cref{eq:bitcalc}.

			\setmyleftlen{clear..fall}
			\begin{align}
				\begin{aligned}
					\beta_{(0,i)}[i] &= 
						\begin{cases}
							HD(\alpha_{(0,i)}[i]) & \quad \text{if } i \in \mathcal{A}\\
							0 & \quad \text{otherwise}
						\end{cases} &&& ,\,\, & i \in [0,N-1]\\\\
						&\backup\begin{aligned}
							\beta_v[i] &= \beta_l[i] \oplus \beta_r[i]\\
							\beta_v\left[i+\frac{\Lambda}{2}\right] &= \beta_r[i]\\		
						\end{aligned}&&& ,\,\, & i \in \left[0,\frac{\Lambda}{2}-1\right]\\
					\end{aligned}
			\label{eq:bitcalc}
			\end{align}

		\subsubsection{Fast simplified successive cancellation}
		\label{sect:fssc}
			The fast simplified successive cancellation (FSSC) decoding algorithm \cite{4} improves upon the computational complexity of the SC decoding algorithm by recognizing constituent codes in the SC decoding tree and pruning the nodes. The four nodes considered are:

			\begin{itemize}
				\item \textbf{Rate-0}

					Rate-0 (R-0) nodes are the nodes in the SC tree below which all the leaf nodes correspond to frozen bits. For an R-0 node at $v=(\lambda,\phi)$ in the decoding tree, no further traversal is needed and the bit estimates for the stage can be update as follows:

					\begin{equation*}
						\beta_v[i] = 0\quad\;,\;i \in [0,\Lambda-1]
					\end{equation*}
				\clearpage
				\item\textbf{Repetition}

					Repetition (REP) nodes contain only a single information bit at the rightmost leaf node. The bit estimates for REP node at $v=(\lambda,\phi)$ in the tree can therefore only be all 0's or all 1's, and the decision is made using an efficient ML decoding by:

					\begin{equation*}
						\beta_v[i] = HD\left(\sum_{k=0}^{\Lambda-1} \alpha_v[k]\right)\quad\;,\;i \in [0,\Lambda-1]
					\end{equation*}		

				\item\textbf{Rate-1}

					Rate-1 (R-1) nodes are the nodes in the SC tree below which all the leaf nodes correspond to information bits. Similar to R-0 nodes, an R-1 node at $v=(\lambda,\phi)$ requires no further traversal and the bit estimates for the stage can be updated by taking a hard decision on the stage LLRs:

					\begin{equation*}
						\beta_v[i] = HD\left(\alpha_v[i]\right)\quad\;,\;i \in [0,\Lambda-1]
					\end{equation*}		

				\item \textbf{Single parity check}

					Single parity check (SPC) nodes contain only a single frozen bit at the leftmost leaf node. The bit estimates for REP node at $v=(\lambda,\phi)$ in the tree therefore have to satisfy a parity constraint such that the XOR of all the estimates should be 0. This can be achieved by computing the parity of the hard decisions of the REP node LLRs, and then flipping the least reliable estimate if the parity is 1:

					\begin{align*}
						parity &= \XORsum_{k=0}^{\Lambda-1} HD\left(\alpha_v[k]\right) \\\\
						j &= \argmin_i \left|\alpha_v[i]\right|\quad\;,\;i \in [0,\Lambda-1]\\\\
						\beta_v[i] &= 
			  			\begin{cases}
			    			HD\left(\alpha_v[i]\right) \oplus parity & \quad \text{if } i=j\\
			    			HD\left(\alpha_v[i]\right) & \quad \text{otherwise }
			  		\end{cases}\quad\;,\;i \in [0,\Lambda-1]
					\end{align*}		

			\end{itemize}

			Since the FSSC scheme does not traverse the decoding tree till the leaf nodes, the bit estimates $\hat{u}_i$ are not readily available. With non-systematic encoding, $\hat{u}_i$ can be obtained by re-encoding the estimated codeword present in the bit estimates at the root node of the tree, $\beta_{(n,0)}[i]$, $i \in [0,N-1]$. With systematic encoding \cite{3,30}, $\hat{u}_i$ is directly available in $\beta_{(n,0)}[i]$, $i \in [0,N-1]$.

		\subsubsection{Successive cancellation list}

			When the SC decoding algorithm encounters an information bit, an immediate decision is made and half the potential remaining paths are discarded from consideration. On the other hand, by considering both possibilities for information bits, ML decoding performance is achieved at the cost of searching through paths that grow exponentially in number. The successive cancellation list (SCL) decoding algorithm \cite{26,5} is a trade-off between these two extremes in that it limits the number of paths under consideration to a fixed list size $L$. At each information bit index, the number of paths is doubled. When the number of paths exceeds $L$, the decoder only considers the $L$ most reliable paths and discards the rest. 

			In order to ascertain which paths should remain in the list and which should be discarded, each path is associated with a path metric (PM) that is updated using the LLRs when a decision is made at the leaf nodes for bit index $i$, as shown in (\ref{eq:PMupd}) \cite{6}.

			\begin{align} \label{eq:PMupd}
				PM_l &= 
					\begin{cases}
			    	PM_l & \quad \text{if } \hat{u}_i=HD(\alpha_{(0,i),\;l})\\
			    	PM_l+|\alpha_{(0,i),\;l}|& \quad \text{otherwise }
			  	\end{cases}&, \forall\;l\text{ paths in the list}
			\end{align}

			After the SCL decoder has estimated all $N$ bits, the path with the best PM is returned as the decoding output. Results in \cite{5} show a significant improvement in error correction performance by appending a small cyclic redundancy check (CRC) code with the message bits to aid the SCL decoder in choosing the correct path from the final candidates in the list. 

		\subsubsection{Fast simplified successive cancellation list} \label{sect:fsscl}

			The FSSC scheme is applied to SCL decoding in \cite{7}, by defining the path creation and PM update for an FSSC node $v$ located at $(\lambda,\phi)$ as follows:

			\begin{itemize}
				\item \textbf{Rate-0}

					An R-0 node creates no new paths, and the PM's and node bit estimates are updated according to:

					\begin{align*}
						\begin{aligned}
							\beta_{v,l}[i] &= 0\quad\;,\;i \in [0,\Lambda-1]\\
							PM_l &= PM_l+\sum_{i=0}^{\Lambda-1}{HD(\alpha_{v,l}[i])\;|\alpha_v[i]|} 
						\end{aligned}&&, \forall\;l\text{ paths in the list}
					\end{align*}

				\item\textbf{Repetition}

					REP nodes create only two candidate paths for each path in the list, and the bit estimates and PM updates are given by:

					\begin{align*}
						\begin{aligned}
							\beta_{v,l}^{(1)}[i] &= 0\quad\;,\;i \in [0,\Lambda-1]\\
							PM_l^{(1)} &= PM_l+\sum_{i=0}^{\Lambda-1}{HD(\alpha_v[i])\;|\alpha_v[i]|}\\\\
							\beta_{v,l}^{(2)}[i] &= 1\quad\;,\;i \in [0,\Lambda-1]\\
							PM_l^{(2)} &= PM_l+\sum_{i=0}^{\Lambda-1}{\left(1-HD(\alpha_v[i])\right)\;|\alpha_v[i]|}
						\end{aligned} &&, \forall\;l\text{ paths in the list}&&\\
					\end{align*}

				\clearpage
				\item\textbf{Rate-1}

					R-1 nodes are limited in the number of paths that are created following the Chase-II decoding algorithm. Each path in the list is extended with the four possible permutations of flipping the hard decisions of the bits at indices $m_1$ and $m_2$, corresponding to the two least reliable LLR's. The $4L$ paths are then pruned back down to $L$.
					
					\begingroup
					\allowdisplaybreaks
					\begin{align*}
						\begin{aligned}
							\beta_{v,l}^{(1)}[i] &= HD(\alpha_{v,l}[i])\;\;,\;i \in [0,\Lambda-1]\\
							PM_l^{(1)} &= PM_l\\\\
							\beta_{v,l}^{(2)}[i] &= 
							\begin{cases}
								HD(\alpha_{v,l}[i]) \oplus 1 & \quad \text{if } i=m_1\\
								HD(\alpha_{v,l}[i]) & \quad \text{otherwise }
							\end{cases}\;\;,\;i \in [0,\Lambda-1]\\
							PM_l^{(2)} &= PM_l+|\alpha_v[m_1]|\\\\
							\beta_{v,l}^{(3)}[i] &= 
							\begin{cases}
								HD(\alpha_{v,l}[i]) \oplus 1 & \quad \text{if } i=m_2\\
								HD(\alpha_{v,l}[i]) & \quad \text{otherwise }
							\end{cases}\;\;,\;i \in [0,\Lambda-1]\\
							PM_l^{(3)} &= PM_l+|\alpha_v[m_2]|\\\\
							\beta_{v,l}^{(4)}[i] &= 
							\begin{cases}
								HD(\alpha_{v,l}[i]) \oplus 1 & \quad \text{if } i\in\{m_1,m_2\}\\
								HD(\alpha_{v,l}[i]) & \quad \text{otherwise }
							\end{cases}\;\;,\;i \in [0,\Lambda-1]\\
							PM_l^{(4)} &= PM_l+|\alpha_v[m_1]|+|\alpha_v[m_2]|\\
						\end{aligned} &&, \forall\;l\text{ paths in the list}&&\\
					\end{align*}
					\endgroup

				\item \textbf{Single parity check}

					SPC nodes are more complex than the preceding nodes discussed because all candidate paths created have to pass the parity check of the node. The number of candidate paths created in an SPC node are limited in a manner similar to R-1 nodes. Upon determining the indices $m_1$, $m_2$, $m_3$ and $m_4$ of the four least reliable LLR's, there are 16 possible permutations of flipping the hard decisions of the bits at these indices, of which, only the half that satisfy the parity constraint are considered. The SPC node thus creates 8 candidate paths from each path in the list, and the total of $8L$ paths are then pruned down to $L$. The bit estimates and PM update equations for the SPC node are omitted for the sake of brevity, and can be referenced from \cite{7}.

			\end{itemize}

		\subsubsection{Successive cancellation stack}

			The SCL decoder considers $L$ candidate paths for each bit estimate in the codeword, resulting in a total search space of $LN$ paths. At this point, the term \textit{iteration} is defined as a decoder making a leaf node bit estimate for a candidate path. The SC decoder therefore takes $N$ iterations to produce the decoding result, while the SCL decoder takes $NL$ iterations. 

			The successive cancellation stack (SCS) algorithm \cite{10} is a sequential traversal through the same search space as the SCL decoder. The algorithm begins by extending an initial path following the SC procedure, and updating its PM following \cref{eq:PMupd}. At the time of estimating information bits, both candidates are considered and the less reliable path is stored in a \textit{stack} of size $D$ that is assumed to be sufficiently large. As the algorithm proceeds, the number of candidates in the stack grows, and in each iteration only the path with the winning PM is extended. 

			The stack contains candidates of different lengths, and over the course of decoding if $L$ paths of length $\Omega \in [1,N]$ have been extended, then all paths with length $\omega \leq \Omega$ are removed from the stack \cite{11}, thus ensuring the same search space as SCL.

			If the winning path has a length of $N$, its bit estimates are returned as the decoding result and the algorithm terminates. Alternatively, the CRC-aided scheme in SCL can be applied to validate the decoded result \cite{11}. If the CRC check fails, the path is removed from the stack and the algorithm continues. By nature of the algorithm, if $L$ paths fail the final CRC check, then all paths are removed from the stack and the algorithm terminates.

			An upper bound on the size of the stack is $D=LN$ \cite{11}, which is the maximum number of paths the SCS algorithm can investigate. While results in \cite{10,11,12,13,14} show that it is possible to achieve similar error correction performance with much smaller values of $D$ (especially at high SNR's), there is no general approach to determining the smaller value of D for different code parameters and channel conditions. In SCS implementations where $D$ is less than the upper bound and the stack is full, new candidate paths replace the path with the least reliable PM, and only if the PM of the new path is more reliable itself.

\section{Implementation Details} \label{s3}

	This section introduces the memory layout and decoding schedule implementation for the successive cancellation family of polar decoders, which is then extended to incorporate list decoding. Finally, the stack decoder implementation is discussed.

	\subsection{Successive cancellation decoders} \label{sect:scimplem}

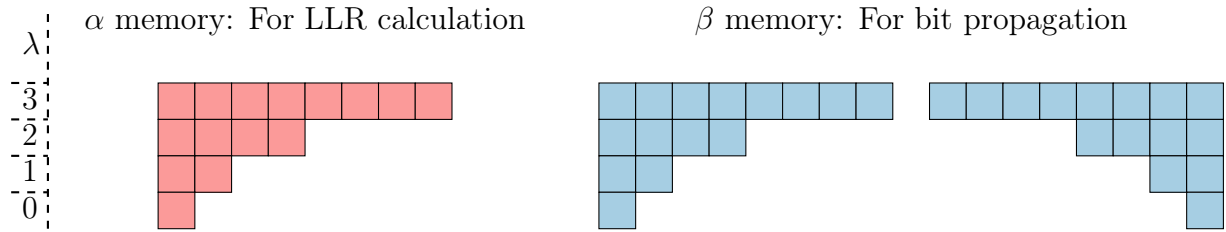
\begin{figure}[t]
			\centering
			\begin{tikzpicture}[scale=\textwidth/33 cm,every node/.style={scale=1}]

				\tikzset{
					a/.style={fill=Paired-6},
					b/.style={fill=Paired-2},
				}

				\draw[thick, dashed] (-5,0) -- (-5,6);
				\draw[thick, dashed] (-6,4) -- (-5,4);
				\draw[thick, dashed] (-6,3) -- (-5,3);
				\draw[thick, dashed] (-6,2) -- (-5,2);
				\draw[thick, dashed] (-6,1) -- (-5,1);

				\node[anchor=south west] at (-6,4.5) {$\lambda$};
				\node[anchor=south west] at (-6,3) {3};
				\node[anchor=south west] at (-6,2) {2};
				\node[anchor=south west] at (-6,1) {1};
				\node[anchor=south west] at (-6,0) {0};

				\node[anchor=south] at (2,5) {$\alpha$ memory: For LLR calculation};
				
				\draw[a] (-2,3) rectangle (-1,4);
				\draw[a] (-1,3) rectangle (0,4);
				\draw[a] (0,3) rectangle (1,4);
				\draw[a] (1,3) rectangle (2,4);
				\draw[a] (2,3) rectangle (3,4);
				\draw[a] (3,3) rectangle (4,4);
				\draw[a] (4,3) rectangle (5,4);
				\draw[a] (5,3) rectangle (6,4);
				\draw[a] (-2,2) rectangle (-1,3);
				\draw[a] (-1,2) rectangle (0,3);
				\draw[a] (0,2) rectangle (1,3);
				\draw[a] (1,2) rectangle (2,3);
				\draw[a] (-2,1) rectangle (-1,2);
				\draw[a] (-1,1) rectangle (0,2);
				\draw[a] (-2,0) rectangle (-1,1);

				\node[anchor=south] at (18.5,5) {$\beta$ memory: For bit propagation};

				\draw[b] (10,3) rectangle (11,4);
				\draw[b] (11,3) rectangle (12,4);
				\draw[b] (12,3) rectangle (13,4);
				\draw[b] (13,3) rectangle (14,4);
				\draw[b] (14,3) rectangle (15,4);
				\draw[b] (15,3) rectangle (16,4);
				\draw[b] (16,3) rectangle (17,4);
				\draw[b] (17,3) rectangle (18,4);
				\draw[b] (10,2) rectangle (11,3);
				\draw[b] (11,2) rectangle (12,3);
				\draw[b] (12,2) rectangle (13,3);
				\draw[b] (13,2) rectangle (14,3);
				\draw[b] (10,1) rectangle (11,2);
				\draw[b] (11,1) rectangle (12,2);
				\draw[b] (10,0) rectangle (11,1);

				\draw[b] (19,3) rectangle (20,4);
				\draw[b] (20,3) rectangle (21,4);
				\draw[b] (21,3) rectangle (22,4);
				\draw[b] (22,3) rectangle (23,4);
				\draw[b] (23,3) rectangle (24,4);
				\draw[b] (24,3) rectangle (25,4);
				\draw[b] (25,3) rectangle (26,4);
				\draw[b] (26,3) rectangle (27,4);
				\draw[b] (23,2) rectangle (24,3);
				\draw[b] (24,2) rectangle (25,3);
				\draw[b] (25,2) rectangle (26,3);
				\draw[b] (26,2) rectangle (27,3);
				\draw[b] (25,1) rectangle (26,2);
				\draw[b] (26,1) rectangle (27,2);
				\draw[b] (26,0) rectangle (27,1);

			\end{tikzpicture}
			\par\bigskip
		\caption{$\alpha$ and $\beta$ memory for $PC(8,4)$.}
		\label{fig:scmems}
		\end{figure}

		The SC and FSSC algorithms make use of the $\alpha$ and a $\beta$ memory tree structures shown in \cref{fig:scmems} to store intermediate LLR calculations and bit propagations. The memory is structured according to the space efficient scheme outlined in \cite{5}, and has a spatial complexity $O(N)$ that scales linearly with the code length $N$. 

		Each stage $\lambda$ in the $\alpha$ memory is only given $\Lambda$ slots of memory - enough to store the LLR's of a single branch $\phi$. This is possible because upon observing the SC schedule, one can see that when calculating the LLR's $\alpha_v[i]$ at a node $v = (\lambda,\phi)$, the LLR's for all branches $\phi' < \phi$ in the same stage $\lambda$ will not be used again and can be safely overwritten.

		Each stage in the $\beta$ memory is given $2\Lambda$ slots of memory. This is because a stage must store the bit estimates from two child branches in order to update the parent node in the stage above. Once the bit estimates have been propagated, the values can be safely overwritten by subsequent nodes in the stage.

		The schedule of operations in the SC decoder is realized at run time using the index $i$ of the current bit $\hat{u}_i$ being estimated, by implementing \cref{eq:llrcalc,eq:bitcalc} according to \cref{alg:reca,alg:recb} \cite{5}.

		\begin{algorithm}[t]
			\label{alg:reca}
			\caption{\function{recursively\_calc\_$\alpha$($\lambda,\phi$)}}
			\BlankLine
			
			\If{$\lambda = n$} {
				\textbf{return\\}
			}
			
			\If{$\phi$ is even} {
				\function{recursively\_calc\_$\alpha$}($\lambda+1, \floor*{\frac{\phi}{2}}$)\\
			}
			
			\BlankLine	
			$v = (\lambda,\phi)$\\
			$u = (\lambda,\,\phi-1)$\\
			$p = \left(\lambda+1,\,\floor*{\frac{\phi}{2}}\right)$\\
			$\Lambda = 2^\lambda$\\
			\BlankLine

			\For{$i = 0,1,\dots,\Lambda-1$} {
				\eIf{$\phi$ is even}{
			  	$\alpha_v[i] = f(\alpha_p[i],\,\alpha_p[i+\Lambda])$\\
			  }{
			  	$\alpha_v[i] = g(\alpha_p[i],\,\alpha_p[i+\Lambda],\,\beta_u[i])$\\
				}
			}
		\end{algorithm}

		\begin{algorithm}[t]
			\label{alg:recb}
			\caption{\function{recursively\_update\_$\beta$}($\lambda,\phi$)}
			\BlankLine
			\If{$\phi$ is even} {
				\textbf{return\\}
			}
			\BlankLine
			$v = (\lambda+1,\,\floor*{\frac{\phi}{2}})$\\
			$l = (\lambda,\,\phi-1)$\\
			$r = (\lambda,\,\phi)$\\
			$\Lambda = 2^\lambda$\\
			\BlankLine
			\For{$i = 0,1,\dots,\Lambda-1$}{
				$\beta_v[i] = \beta_l[i] \oplus \beta_r[i]$\\
				$\beta_v\left[i+\Lambda\right] = \beta_r[i]$\\		
			}
			
			\function{recursively\_update\_$\beta$}($\lambda+1, \floor*{\frac{\phi}{2}}$)\\
		\end{algorithm}

		Computing the FSSC schedule at run time incurs a significant computational penalty since the entire decoding tree has to be traversed to identity the FSSC nodes. The FSSC schedule is therefore created and stored as the decoder is instantiated, which the decoder can then load and loop through for each decoding run. The schedule is stored as operations and the nodes in the tree at which they are performed.  \Cref{fig:fsscsch} illustrates an example of an FSSC schedule created for the SC decoding tree in \cref{fig:scdectree}. By abuse of notation, the operations that implement \cref{eq:llrcalc,eq:bitcalc} are denoted by $\boldsymbol{\alpha}$ and $\boldsymbol{\beta}$ respectively, and the operations \textbf{R-0}, \textbf{R-1}, \textbf{REP} and \textbf{SPC} implement the equation for the corresponding node.

		\begin{figure}[t]
			\centering
			\begin{tikzpicture}
				\node[anchor=east] at (0,1) {Operation:};
				\node[anchor=east] at (0,0) {Node:     };

				\node[anchor=east] at (2,1) {$\boldsymbol{\alpha}$};
				\node[anchor=east] at (2,0) {(2,0)     };

				\node[anchor=east] at (4,1) {\textbf{REP}};
				\node[anchor=east] at (4,0) {(2,0)     };

				\node[anchor=east] at (6,1) {$\boldsymbol{\alpha}$};
				\node[anchor=east] at (6,0) {(2,1)     };

				\node[anchor=east] at (8,1) {\textbf{SPC}};
				\node[anchor=east] at (8,0) {(2,1)     };

				\node[anchor=east] at (10,1) {$\boldsymbol{\beta}$};
				\node[anchor=east] at (10,0) {(2,1)     };

			\end{tikzpicture}
		\caption{FSSC schedule for $PC(8,4)$ with $\mathcal{A} = {3,5,6,7}$.}
		\label{fig:fsscsch}
		\end{figure}

	\subsection{List decoders} \label{sect:sclimp}

		The SCL family of decoders extend up to $L$ paths simultaneously, each of which have different values for intermediate LLR's and bit estimates. The SCL and FSSCL therefore instantiate $L$ copies of the $\alpha$ and a $\beta$ memory trees from \cref{sect:scimplem}, resulting in a spatial complexity of $O(LN)$.

		The naive approach to use these memory trees is to duplicate the $\alpha$ and $\beta$ values for new candidate paths, which results in wasted memory operations for paths that are killed before the values are used. 

		The authors in \cite{5} propose a \textit{lazy-copy} scheme in which $\alpha$ and $\beta$ memory is allocated stage by stage, rather than the tree as a whole, and new candidate paths point to the memory of the parent path that created them. Memory duplication now only occurs when a path needs to modify a stage in memory pointed to by multiple paths, and only that stage is duplicated. The SCL and FSSCL decoders in this work implement a minor modification to the \textit{lazy-copy} scheme of \cite{5} to support decoding in the LLR domain.

		The decoding schedule for SCL and FSSCL is realized in the same manner as outlined for their counterparts SC and FSSC.

	\subsection{Stack decoder}

		A candidate path that is placed on the stack must store its: 

		\begin{itemize}
			\item path metric (PM)
			\item path length (PL)
			\item bit estimates $\hat{u}_i\,,\,i \in [0,N-1]$
			\item intermediate $\alpha$ and $\beta$ values
		\end{itemize}

		The stack is implemented as $D$ length arrays of these data-structures, and when a path is placed on the stack, it is assigned an index at which to store its values in these arrays. The winning path for each iteration is determined through a linear search on the PM arrays.

		The PM and PL arrays are one-dimensional with a space complexity of $O(D)$, while the bit estimates array is two dimensional with a complexity of $O(DN)$. The data-structure for the $\alpha$ and $\beta$ values follows the same structure as in \cref{fig:scmems}, resulting in a memory complexity of $O(DN)$. Its usage is also governed by the \textit{lazy copy} scheme in \cite{5}. Finally, the schedule for SCS is realized following the same \cref{alg:reca,alg:recb} as in the SC decoder.

		Based on the observation that the SCS decoder extends only one path at a time, a reduced memory scheme (SCS-RM) is proposed in \cite{22} in which only a single copy of the $\alpha$ and $\beta$ memory is instantiated. The initial path is created, and as long as there is no path switch, intermediate $\alpha$ and $\beta$ values remain valid and the path can continue to be extended. Potential candidates that are created store only their PM, PL and leaf node bit estimates $\hat{u}_0^{i-1}$, where $i$ is the current length of the path.

		A path switch renders the $\alpha$ and $\beta$ memory values invalid, which now have to be re-calculated for the new path. This is achieved by first populating the $\beta$ memory with the estimates $\hat{u}_0^{i-1}$, following which the $\alpha$ memory is updated by initiating the calculation of $LLR\left(y_0^{N-1}, \hat{u}_0^{i-1}| \hat{u}^i\right)$ from the channel values at the root node of the decoding tree, rather than from an intermediate stage as dictated by the standard SC decoding procedure. The $\alpha$ and $\beta$ memory can now be used following the traditional SC schedule until the next path switch is encountered, at which point the re-calculation is performed again

		Populating the $\beta$ memory for the newly switched path $p$ uses the same \cref{alg:recb} defined for the SC schedule. The procedure is highlighted in \cref{alg:ldp}, which is performed only once when the path is switched. Recalculating the $\alpha$ memory from the root node requires a minor modification to the SC \cref{alg:reca}, as highlighted in \cref{alg:calp}.

		\begin{algorithm}[t]
			\label{alg:ldp}
			\caption{\function{Populating $\beta$ memory with a new path $p$}}
			\BlankLine

			\For{$i=0,1,\dots,PL_p-1$} {
				$\beta_{(0,i)}[0] = \hat{u}_p[i]$\\
				\function{recursively\_update\_$\beta$}($0, i$)\\

			}
		\end{algorithm}

		\begin{algorithm}[t]
			\label{alg:calp}
			\caption{\function{recursively\_calc\_$\alpha$($\lambda,\phi$) {\color{red} modified for SCS-RM}}}
			\BlankLine
			
			\If{$\lambda = n$} {
				\textbf{return\\}
			}
			
			\If{$\phi$ is even {\color{red}{\upshape \textbf{or}} path has switched}} {
				\function{recursively\_calc\_$\alpha$}($\lambda+1, \psi$)\\
			}
			
			\BlankLine	
			$v = (\lambda,\phi)$\\
			$u = (\lambda,\,\phi-1)$\\
			$p = \left(\lambda+1,\,\floor*{\frac{\phi}{2}}\right)$\\
			$\Lambda = 2^\lambda$\\
			\BlankLine

			\For{$\beta = 0,1,\dots,\Lambda$} {
				\eIf{$\phi$ is even}{
			  	$\alpha_v[i] = f(\alpha_p[i],\,\alpha_p[i+\Lambda])$\\
			  }{
			  	$\alpha_v[i] = g(\alpha_p[i],\,\alpha_p[i+\Lambda],\,\beta_u[i])$\\
				}
			}
		\end{algorithm}

\section{Fast simplified stack decoding} \label{s4}

	The FSSCL scheme of \cite{7} can readily be applied to the SCS decoder. The key difference is that the FSSCL decoder has all candidate paths available at a given node, and is able to prune paths and pick the survivors immediately. In contrast, the FSSCS decoder creates all the candidate paths for the node and places them on the stack, and the paths are either further extended or killed at a later point following the SCS algorithm rules.

	Two key implementational details are highlighted, the first of which is that the FSSCS decoder switches between paths at different points in the FSSC schedule. While the path length alone can be used to determine the coordinates of the current node $(\lambda,\phi)$ in the decoding tree, it is not sufficient to determine which FSSC operation ($\boldsymbol{\alpha}$, $\boldsymbol{\beta}$, \textbf{R-0}, \textbf{R-1}, \textbf{REP} or \textbf{SPC}) must be performed. To this end, when a path is placed on the stack, it stores and additional parameter - its current progress in the FSSC schedule.

	The second detail involves applying the SCS-RM scheme of \cite{22} to the FSSCS decoder, referred to as FSSCS-RM. Since the FSSC scheme does not necessarily traverse down to the root nodes to make bit estimates, it is impossible for the FSSCS-RM decoder to repopulate the $\beta$ memory via the SCS-RM \cref{alg:ldp}. This hurdle is overcome by changing the structure of the $\beta$ memory of the FSSCS-RM decoder. The work in \cite{27} presents an efficient scheme to compute and store the $\beta$ values in the context of VLSI design, which is adapted to software in this work.

\begin{figure}[t]
		\begin{subfigure}{0.5\textwidth}
			\begin{tikzpicture}[scale=\textwidth/17 cm]
				\tikzset{
					b/.style={fill=Paired-2},
					n/.style={anchor=west},
				}

				\draw[b] (0 ,0) rectangle (2 ,2);
				\draw[b] (2 ,0) rectangle (4 ,2);
				\draw[b] (4 ,0) rectangle (6 ,2);
				\draw[b] (6 ,0) rectangle (8 ,2);
				\draw[b] (8 ,0) rectangle (10,2);
				\draw[b] (10,0) rectangle (12,2);
				\draw[b] (12,0) rectangle (14,2);
				\draw[b] (14,0) rectangle (16,2);
			\end{tikzpicture}
		\caption{$\beta$ memory structure.}
		\label{fig:fsscsbmema}	
		\end{subfigure}%
		\begin{subfigure}{0.5\textwidth}
			\begin{tikzpicture}[scale=\textwidth/17 cm]
				\tikzset{
					b/.style={fill=Paired-2},
					o/.style={scale=0.9,anchor=west},
					n/.style={red,o},
				}

				\draw[b] (0 ,0) rectangle (2 ,2);
				\draw[b] (2 ,0) rectangle (4 ,2);
				\draw[b] (4 ,0) rectangle (6 ,2);
				\draw[b] (6 ,0) rectangle (8 ,2);
				\draw[b] (8 ,0) rectangle (10,2);
				\draw[b] (10,0) rectangle (12,2);
				\draw[b] (12,0) rectangle (14,2);
				\draw[b] (14,0) rectangle (16,2);

				\node[n] at (0,1) {$\beta_{(2,0)}^{[0]}$};
				\node[n] at (2,1) {$\beta_{(2,0)}^{[1]}$};
				\node[n] at (4,1) {$\beta_{(2,0)}^{[2]}$};
				\node[n] at (6,1) {$\beta_{(2,0)}^{[3]}$};
			\end{tikzpicture}
		\caption{$\beta$ memory contents following \textbf{REP} node.}
		\label{fig:fsscsbmemb}	
		\end{subfigure}
		\par\bigskip
		\begin{subfigure}{0.5\textwidth}
			\begin{tikzpicture}[scale=\textwidth/17 cm]
				\tikzset{
					b/.style={fill=Paired-2},
					o/.style={scale=0.9,anchor=west},
					n/.style={red,o},
				}

				\draw[b] (0 ,0) rectangle (2 ,2);
				\draw[b] (2 ,0) rectangle (4 ,2);
				\draw[b] (4 ,0) rectangle (6 ,2);
				\draw[b] (6 ,0) rectangle (8 ,2);
				\draw[b] (8 ,0) rectangle (10,2);
				\draw[b] (10,0) rectangle (12,2);
				\draw[b] (12,0) rectangle (14,2);
				\draw[b] (14,0) rectangle (16,2);

				\node[o] at (0 ,1) {$\beta_{(2,0)}^{[0]}$};
				\node[o] at (2 ,1) {$\beta_{(2,0)}^{[1]}$};
				\node[o] at (4 ,1) {$\beta_{(2,0)}^{[2]}$};
				\node[o] at (6 ,1) {$\beta_{(2,0)}^{[3]}$};
				\node[n] at (8 ,1) {$\beta_{(2,1)}^{[0]}$};
				\node[n] at (10,1) {$\beta_{(2,1)}^{[1]}$};
				\node[n] at (12,1) {$\beta_{(2,1)}^{[2]}$};
				\node[n] at (14,1) {$\beta_{(2,1)}^{[3]}$};
			\end{tikzpicture}
		\caption{$\beta$ memory contents following \textbf{SPC} node.}
		\label{fig:fsscsbmemc}	
		\end{subfigure}%
		\begin{subfigure}{0.5\textwidth}
			\begin{tikzpicture}[scale=\textwidth/17 cm]
				\tikzset{
					b/.style={fill=Paired-2},
					o/.style={scale=0.9,anchor=west},
					n/.style={red,o},
				}

				\draw[b] (0 ,0) rectangle (2 ,2);
				\draw[b] (2 ,0) rectangle (4 ,2);
				\draw[b] (4 ,0) rectangle (6 ,2);
				\draw[b] (6 ,0) rectangle (8 ,2);
				\draw[b] (8 ,0) rectangle (10,2);
				\draw[b] (10,0) rectangle (12,2);
				\draw[b] (12,0) rectangle (14,2);
				\draw[b] (14,0) rectangle (16,2);

				\node[n] at (0 ,1) {$\beta_{(3,0)}^{[0]}$};
				\node[n] at (2 ,1) {$\beta_{(3,0)}^{[1]}$};
				\node[n] at (4 ,1) {$\beta_{(3,0)}^{[2]}$};
				\node[n] at (6 ,1) {$\beta_{(3,0)}^{[3]}$};
				\node[n] at (8 ,1) {$\beta_{(3,0)}^{[4]}$};
				\node[n] at (10,1) {$\beta_{(3,0)}^{[5]}$};
				\node[n] at (12,1) {$\beta_{(3,0)}^{[6]}$};
				\node[n] at (14,1) {$\beta_{(3,0)}^{[7]}$};
			\end{tikzpicture}
		\caption{$\beta$ memory contents following propagation.}
		\label{fig:fsscsbmemd}	
		\end{subfigure}
	\caption{$\beta$ memory structure and usage in the FSSCS-RM decoder for $PC(8,4)$.}
	\label{fig:fsscsbmem}
	\end{figure}
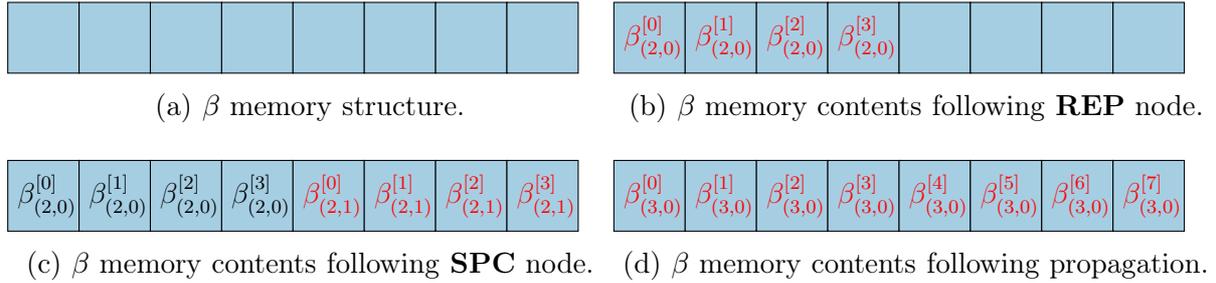

	The $\beta$ memory is now organized as an array of $N$ bits as shown in \cref{fig:fsscsbmema}. When a nodes bit estimates are made following the FSSCL equations, the estimates are stored directly in the $\beta$ memory array beginning at the index corresponding to the length of the path. In the case of a $\boldsymbol{\beta}$ propagation operation, the bits are XOR-ed in place. \Cref{fig:fsscsbmem} shows the usage of the $\beta$ memory for the FSSC schedule of \cref{fig:fsscsch}. In \cref{fig:fsscsbmemb,fig:fsscsbmemc}, the four bits corresponding to the REP and SPC node respectively are stored at the correct locations, following which \cref{fig:fsscsbmemd} shows the $\boldsymbol{\beta}$ operation performed in place. 

	The final content of the $\beta$ memory is the estimated codeword at the root node of the tree, and by using systematic encoding, the estimated message bits are readily available. This leads to the observation that each path on the stack can store the $\beta$ array directly instead of the bit estimates $\hat{u}_i$. An additional advantage is that a path switch in the FSSCS-RM scheme does not need to re-populate the $\beta$ array, since the propagations are already correctly stored in place.

\section{Results and discussion} \label{s5}

	Simulations are performed for $PC(1024,512)$, and the set of information bit indices $\mathcal{A}$ is obtained from the polar code sequence listed in the 3GPP technical specification for the 5G standard \cite{e1}. The CRC used in all variants of the SCL and SCS decoders is the 24-bit CRC-24C with a polynomial of 0xB2B117, also provided in \cite{e1}. The list parameter $L$ is set to 8 and the stack size $D$ is set to the maximum size $NL = 8192$ for all decoders.

	All code is written in C language and compiled with GCC version 6.3.0 using the \texttt{-Ofast, -march=native, -funroll-loops} and \texttt{-finline-functions} compile flags. $\alpha$ and $\beta$ values are implemented using 32-bit floating point numbers and 8-bit unsigned integers respectively. Simulations are run using 6 threads on an AMD Ryzen 5 1600 6-Core CPU clocked at 3.2 GHz. The T/P of the decoder is reported as \textbf{an average per thread}, and considering only information bits.

\begin{figure}[t]
		\centering
		\begin{subfigure}{\textwidth}
			\centering
			\begin{tikzpicture}[scale=0.9]
				\tikzset{
		   		scl/.style={Paired-13, thick, mark=square},
		   		scs/.style={Paired-5, thick, mark=otimes},
		   		scs-rm/.style={Paired-9, thick, mark=triangle},
		   		fsscl/.style={Paired-1, thick, mark=x},
		   		fsscs/.style={Paired-3, thick, mark=o},
				}
				\begin{semilogyaxis}[
					xlabel=$E_b/N_o$ (dB),
					ylabel=FER,
					scale only axis,
					height= 6 cm,
					width=0.6\textwidth,
					legend style={at={(1,0)},anchor=south west},
					legend cell align={left},
					legend style={nodes={scale=0.7, transform shape}},
					grid=both]
					\addplot[scs] plot coordinates {
						(0.00000e+00, 9.83108e-01)
						(5.00000e-01, 8.08896e-01)
						(1.00000e+00, 3.65517e-01)
						(1.50000e+00, 5.99700e-02)
						(2.00000e+00, 3.11643e-03)
						(2.50000e+00, 7.90473e-05)
						};
					\addlegendentry{SCS}
					\addplot[scs-rm] plot coordinates {
						(0.00000e+00, 9.83308e-01)
						(5.00000e-01, 8.09095e-01)
						(1.00000e+00, 3.70415e-01)
						(1.50000e+00, 5.77711e-02)
						(2.00000e+00, 3.21647e-03)
						(2.50000e+00, 7.94745e-05)
						};
					\addlegendentry{SCS-RM}
					\addplot[scl] plot coordinates {
						(0.00000e+00, 9.83108e-01)
						(5.00000e-01, 8.09695e-01)
						(1.00000e+00, 3.65917e-01)
						(1.50000e+00, 6.11694e-02)
						(2.00000e+00, 3.10001e-03)
						(2.50000e+00, 7.89473e-05)
						};
					\addlegendentry{SCL}
					\addplot[fsscs] plot coordinates {
						(0.00000e+00, 9.81409e-01)
						(5.00000e-01, 8.10995e-01)
						(1.00000e+00, 3.54223e-01)
						(1.50000e+00, 6.07696e-02)
						(2.00000e+00, 3.68297e-03)
						(2.50000e+00, 1.09841e-04)
						};
					\addlegendentry{FSSCS-RM}
					\addplot[fsscl] plot coordinates {
						(0.00000e+00, 9.82609e-01)
						(5.00000e-01, 8.13293e-01)
						(1.00000e+00, 3.56122e-01)
						(1.50000e+00, 6.18691e-02)
						(2.00000e+00, 4.42263e-03)
						(2.50000e+00, 1.14303e-04)
						};
					\addlegendentry{FSSCL}
				\end{semilogyaxis}
				\tkzDefPoint(0,0){lim}
			\end{tikzpicture}
		\caption{FER.}
		\label{fig:FSSCSa}
		\end{subfigure}
		\par\bigskip\par\bigskip\par\bigskip
		\begin{subfigure}{\textwidth}
			\centering
			\begin{tikzpicture}[scale=0.9]
				\tikzset{
		   		scl/.style={Paired-13, thick, mark=square},
		   		scs/.style={Paired-5, thick, mark=otimes},
		   		scs-rm/.style={Paired-9, thick, mark=triangle},
		   		fsscl/.style={Paired-1, thick, mark=x},
		   		fsscs/.style={Paired-3, thick, mark=o},
				}
				\tkzDefPoint(0,6.9){lim}
				\begin{semilogyaxis}[
					xlabel=$E_b/N_o$ (dB),
					ylabel=T/P (Mb/s),
					scale only axis,
					height= 6 cm,
					width=0.6\textwidth,
					legend style={at={(1,0)},anchor=south west},
					legend cell align={left},
					legend style={nodes={scale=0.7, transform shape}},
					grid=both]
					\addplot[scs] plot coordinates {
						(0.00000e+00, 8.43964e-04)
						(5.00000e-01, 9.68978e-04)
						(1.00000e+00, 1.72932e-03)
						(1.50000e+00, 4.39395e-03)
						(2.00000e+00, 6.74342e-03)
						(2.50000e+00, 6.91506e-03)
						(3.00000e+00, 8.89654e-03)
						};
					\addlegendentry{SCS}
					\addplot[scs-rm] plot coordinates {
						(0.00000e+00, 2.35459e-02)
						(5.00000e-01, 2.58788e-02)
						(1.00000e+00, 3.46253e-02)
						(1.50000e+00, 5.18958e-02)
						(2.00000e+00, 7.21866e-02)
						(2.50000e+00, 1.10722e-01)
						(3.00000e+00, 2.32370e-01)
						};
					\addlegendentry{SCS-RM}
					\addplot[scl] plot coordinates {
						(0.00000e+00, 2.95592e-01)
						(5.00000e-01, 2.96015e-01)
						(1.00000e+00, 2.90247e-01)
						(1.50000e+00, 2.80185e-01)
						(2.00000e+00, 2.90871e-01)
						(2.50000e+00, 3.09693e-01)
						(3.00000e+00, 3.14545e-01)
						};
					\addlegendentry{SCL}
					\addplot[fsscs] plot coordinates {
						(0.00000e+00, 1.69298e-01)
						(5.00000e-01, 1.82393e-01)
						(1.00000e+00, 2.32408e-01)
						(1.50000e+00, 3.42818e-01)
						(2.00000e+00, 4.48689e-01)
						(2.50000e+00, 6.19550e-01)
						(3.00000e+00, 9.25226e-01)
						};
					\addlegendentry{FSSCS-RM}
					\addplot[fsscl] plot coordinates {
						(0.00000e+00, 1.11897e+00)
						(5.00000e-01, 1.15324e+00)
						(1.00000e+00, 1.16887e+00)
						(1.50000e+00, 1.18691e+00)
						(2.00000e+00, 1.20527e+00)
						(2.50000e+00, 1.21292e+00)
						(3.00000e+00, 1.22102e+00)
						};
					\addlegendentry{FSSCL}
				\end{semilogyaxis}
			\end{tikzpicture}
		\caption{T/P.}
		\label{fig:FSSCSb}
		\end{subfigure}
	\caption{Performance of the FSSCS decoder.}
	\label{fig:FSSCS}
	\end{figure}
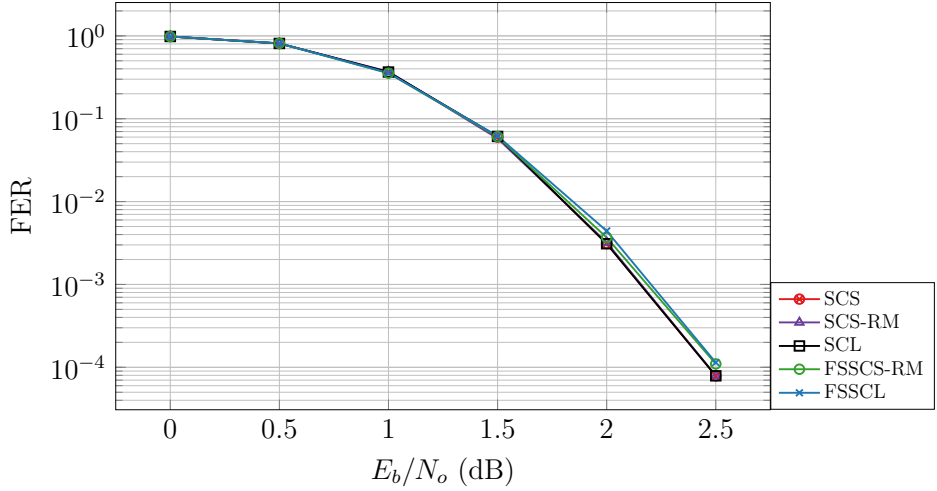
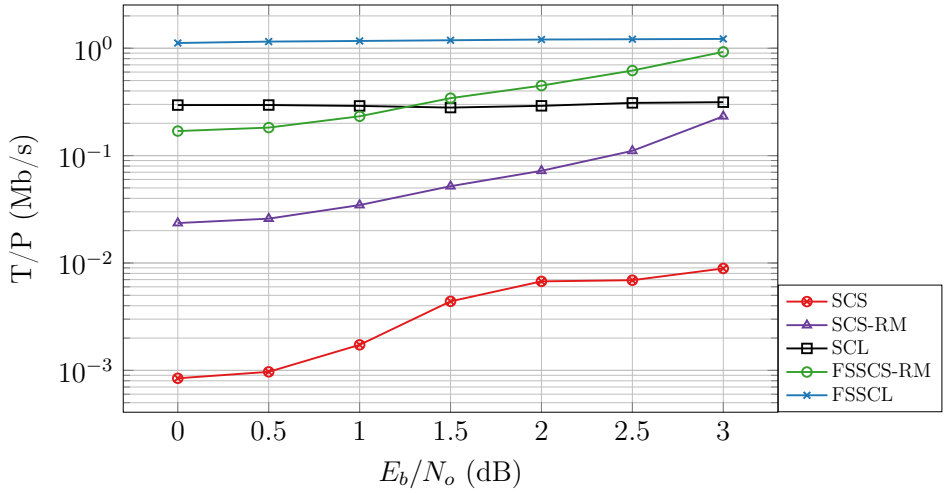

	\Cref{fig:FSSCSa} exhibits that the FER performance is maintained for all variants of the stack and list deocders. The slight FER permornace degradation in the fast simplified decoders is attributed to the Chase-II approximation used \cite{7}. 

	\cref{fig:FSSCSb} shows that the baseline T/P of the SCS decoder is, at best, 9 Kbps at an $E_bN_o$ of 3 dB, which is more than an order of magnitude lower than the SCL T/P of 314 Kbps. The SCS-RM scheme is able to improve the SCS throughput by more than an order of magnitude to 232 Kbps. The FSSCL decoder reports a T/P of 1.22 Mbps, which is four times the T/P of SCL. Finally, applying the fast simplified scheme to SCS decoding results in similar throughput gains as observed with SCL. At an $E_bN_o$ of 3 dB, FSSCS-RM provides a T/P of 930 Kbps, which is four times the T/P of SCS-RM and two orders of magnitude more than the T/P of the baseline SCS.

\section{Conclusion} \label{s6}
	
	This report outlines a procedure for applying the fast simplified scheme \cite{7} to the reduced memory stack decoder \cite{22}. Results show that the T/P of the FSSCS-RM decoder is improved by two orders of magnitude over the baseline SCS decoder, from 9 Kbps to 930 Kbps. The FSSCS-RM decoder using the largest stack size achieves the T/P of the FSSCL decoder at practical SNR's.

\backmatter
\bibliographystyle{ieeetr}
\bibliography{Bibliography}

\end{document}